# Human Leg Training Machine Based on The Multi-linkage System

Yunfei Liu and Zhanghao Yang

**Abstract** —*In real life, many people have leg defects. the goal of our work is to design a mechanism which could help them walk based on a specific trajectory and realize flexible walking finally. In this paper, we use a motor to drive a multi-link leg mechanism. The major issues addressed in this paper are as follows: (i) design human leg training mechanism based on the multi-link mechanism (ii) Simulate leg movement trajectory of multi-link mechanism based on walking process (iii) make use of one motor torque control to control the trajectory and velocity of this mechanism.*

## Introduction

Leg-type robots are currently the mainstream topic. Through simulation, people have explored leg-shaped robots, such as robotic dogs that can move autonomously. In control field, since Iterative Learning Control (ILC) (see [5]-[7]) is a control strategy designed for systems that operate in a repetitive manner over a fixed time interval, thus combining ILC with the design of a motor-driven multi-link mechanism for leg rehabilitation provides significant advancements in the effectiveness and adaptability of rehabilitation robots. People gradually explored the application in many fields with mechanical legs. However, there are currently few robots in the market that help people walk.

Rehabilitation products on the market today are mainly limited to the body's own movement (see [1]-[4]). The principle is mainly to achieve rehabilitation by using the movement of patients' own muscles, which has considerable limitations. Besides, most products are push-off mode, which means the device does not fully simulate the movement of the human leg. At best, they are leg muscle training devices. So, we want to design a motor-driven multi-link mechanism to achieve leg rehabilitation.

In this paper, we design a multi-link mechanism driven by a single motor. The mechanism will simulate the process of human walking and will present a periodic motion. The main advantage of this apparatus is that this mechanism can achieve the perfect cooperation between the legs and the mechanism, and the mechanism plays a standardized auxiliary role so that the human legs can walk follow a specific trajectory. The mechanism is comprised of seven connected rods and driven by a single motor to realize the leg movement of the human leg. By controlling the motor with specific torque, the mechanism can ensure a periodic change of velocity and angle, making sure the start velocity and end velocity to be same, which can simulate the walking mode successfully. This device can be fixed to the waist of the human body in a specific manner and adjusted according to the waist size of the human body. The length of the connecting rod can be adjusted in equal proportion according to the length of the human leg, but the scale between rods and the rod is fixed. The mechanism can fix the ankle, knee and waist, and realize the perfect fusion of human leg and mechanism. For right now, this mechanism is established on the theoretical stage. In the future, we would like to prove the flexibility of human leg joints through a specific manner.

By using the device, people can carry out the training of walk under the action of device assistance. People can control the number of cycles per training, and after a long cycle of auxiliary walking, people can return to a good manner of walking.

## Design

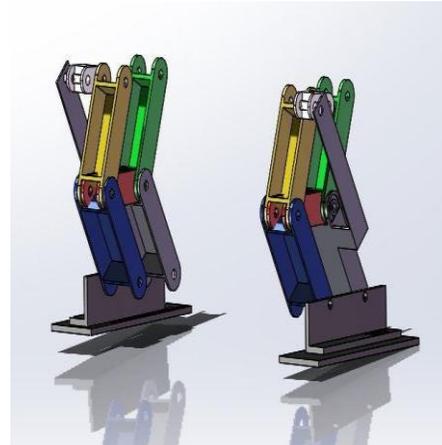

Figure 1. 3D CAD Model

The structure considers the structure of the human body. The left and right independent multi-rod structure can finally be fixed to the waist of the person in a specific manner. There is a specific motor installed on the white link to provide torque on the whole mechanism. In the figure, the white link is the prime rod which will be driven by the motor with a specific torque and the others are the follower links. There is a circular space in the middle of the red rod. The knee can be fixed in a certain way to ensure the flexibility of the knee.

For right now, this mechanism is established on the theoretical stage. In the near future, we would like to prove the flexibility of each joint.

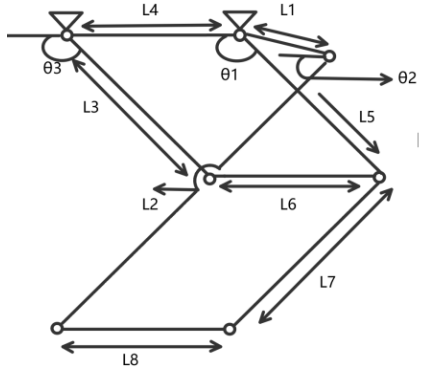

Figure 2. 2D CAD Model

For this mechanism,
Link 1 - Prime Link
Link 2 - Intermediate extension link
Link 4 - Static Link
Link 3, 5, 6, 7, 8 - follower Link

For here, Link 1 represents the prime link to drive the whole mechanism and all the other links are the follower link except static link 4. In order to evaluate the moving trajectory of the mechanism, we set three angles here. However, angle 2 and angle 3 will change with the change of angle 1, and the aim of setting angle 2 and angle 3 is for calculating easier.

In this structure, the ratio between the rods is fixed, following the ratio below:
Link 1: Link 2: Link 4: link 3, 5, 6, 7, 8 = 1: 5: 2: 2.5.

The mechanism rod structure can be adjusted in proportion according to the length of the human leg. Here, we assume a human leg length and use this to set the length of the current links and perform the following simulation and control calculations. The actual length of links is shown below:
Link 1: Link 2: Link 4: link 3, 5, 6, 7, 8 = 18cm: 90cm: 36cm: 45cm.

## Calculation and Assumption

To derive the Lagrange equations the following assumptions were made:
- All joints are revolute.
- Ignore frictional forces at the revolute joints.
- Ignore the mass of link 4 for it will not move.
- Assume the center of gravity is at the geometric center of each link.
- Assume the leg of the patient increases the mass of link 6 and link 8.

The variables in Table I are used to derive the Lagrange equation.

TABLE I

| Symbol | Definition |
|---|---|
| $g$ | Gravitational Acceleration |
| $m_i$ | Mass of the Links (i=1,2,3,4,5,6,7,8) |
| $l_i$ | Length of the Links (i=1,2,3,4,5,6,7,8) |
| $q_i$ | Angle of Links (i=1,2,3) |
| $q_d$ | Desired Angle of Links (i=1,2,3) |
| $\dot{q}(w)$ | Angle Velocity of Links (i=1,2,3) |
| $\dot{q}_d$ | Desired Angle Velocity of Links (i=1,2,3) |
| $\ddot{q}(a)$ | Angle Acceleration of Links (i=1,2,3) |
| $\ddot{q}_d$ | Desired Angle Acceleration of Links (i=1,2,3) |
| $\tau$ | Torque Applying on Link 1 |

As shown in figure 2, this training machine has eight links (seven moving links). The design of this system only introduces three angle variables (q1, q2 and q3), because the angle1 has position, velocity and acceleration limitations on the angle 2 and angle 3, the degree of freedom of this mechanism is reduced to 1, so this is a closed-chain 1-DOF mechanism.

Below is the relationship between angle 1, angle 2 and angle 3.

Position:
$$A = 2l_1 l_2 \sin(q_1) \tag{1}$$
$$B = 2l_2(l_1 \cos(q_1) - l_4) \tag{2}$$
$$C = l_1^2 + l_2^2 + l_4^2 - l_3^2 - 2l_1 l_4 \cos(q_1) \tag{3}$$
$$q_2 = 2\tan^{-1}\left(\frac{A-(A^2+B^2-C^2)^{0.5}}{B-C}\right) \tag{4}$$
$$D = 2l_1 l_3 \sin(q_1) \tag{5}$$
$$E = 2l_3(l_1 \cos(q_1) - l_4) \tag{6}$$
$$F = l_2^2 - l_1^2 - l_3^2 - l_4^2 + 2l_1 l_4 \cos(q_1) \tag{7}$$
$$q_3 = 2\tan^{-1}\left(\frac{D-(D^2+E^2-F^2)^{0.5}}{E-F}\right) \tag{8}$$

Velocity:
$$w_2 = \frac{-w_1 l_1 \sin(q_1-q_3)}{l_2 \sin(q_2-q_3)} \tag{9}$$
$$w_3 = \frac{w_1 l_1 \sin(q_1-q_2)}{l_3 \sin(q_3-q_2)} \tag{10}$$

Acceleration:
$$a_2 = \frac{-w_1^2 l_1 \cos(q_1-q_3) - w_2^2 l_2 \cos(q_2-q_3) + w_3^2 l_3}{l_2 \sin(q_2-q_3)} \tag{11}$$
$$a_3 = \frac{w_1^2 l_1 \cos(q_1-q_2) - w_3^2 l_3 \cos(q_3-q_2) + w_2^2 l_2}{l_3 \sin(q_3-q_2)} \tag{12}$$

The Lagrange equation we need to use is shown below:

$$A(q)\ddot{q} + h(q,\dot{q})\dot{q} + g(q) = \tau \quad (13)$$

Where $A(q)$ is a $3 \times 3$ matrix, $h(q,\dot{q})$ is a $3 \times 3$ matrix, $g(q)$ is a $3 \times 1$ matrix, form this equation we can get the torque 1 that motor needs to generate.

## Controller Design

In order to use a reasonable control method to control the trajectory of the machine, we set the parameters of a sextic polynomial trajectory:

$$q_d = 0.7752 + \frac{5}{2}\pi t^3 - \frac{15}{8}\pi t^4 + \frac{3}{8}\pi t^5 \quad (14)$$

$$\dot{q}_d = \frac{15}{2}\pi t^2 - \frac{15}{2}\pi t^3 + \frac{15}{8}\pi t^4 \quad (15)$$

$$\ddot{q}_d = 15\pi t - \frac{45}{2}\pi t^2 + \frac{15}{2}\pi t^3 \quad (16)$$

Through this motion, the angle 1 starts from 0.7752 rad, 0 rad/s and 0 rad/s², ending with desired 7.0582 rad, 0 rad/s and 0 rad/s², followed by angle 2 and angle 3 with desired motion.

Then we used a linear control law – computed torque control method to track the above trajectory that is equivalent to the feedback scheme equation derived as below:

$$\ddot{q} = v = A^{-1}[\tau g(q) - h(q,\dot{q})] \quad (17)$$
$$v = \ddot{q}_d + k_v(\dot{q}_d - \dot{q}) + k_p(q_d - q) \quad (18)$$
$$\ddot{e} + k_v\dot{e} + k_p e = 0 \quad (19)$$

Where $e = q_d - q$.

## Simulation

To verify the design can reach desired kinematic states in all configurations, equations above were simulated in MATLAB.

The mass of link 6 was simulated as 1kg because it was assumed that it would also bear the weight of the leg. Average weight for leg is 30% of the total weight of body, which is about 20kg based on the mean collected from NCD-RisC (a British medical research institution) [8]. The mass of link 8 was simulated to have 1kg.

The length of links 3 and 5 are the same at 0.45m, which is the mean thigh length for adults based on report from NCD-RisC (a British medical research institution) [8]. The length of link 7 is 0.45m, which is also the average of a human shank.

With the corresponding masses and lengths, the mechanism was then simulated to move q1 from 0.7752 to 7.0582. This simulates the range of motion a human would move his leg in walking, starting at his side, the thigh of leg would move forward and backward, reaching the left boundary and right boundary, while the knee (point B) would bend in a certain degree and back to the original angle as the shank of leg and ankle (point A) move. The simulated range of motion depicts that the machine supported leg exoskeleton can track a desired moving trajectory through each configuration in Fig 4.

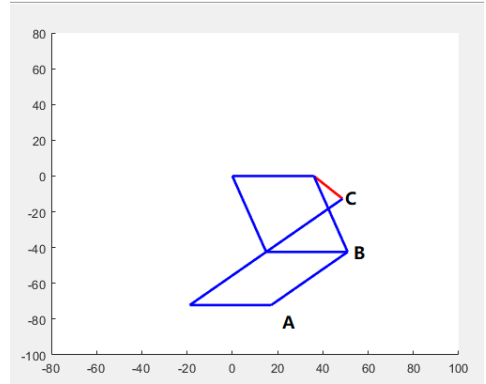

Figure 3. Linkage System Animation

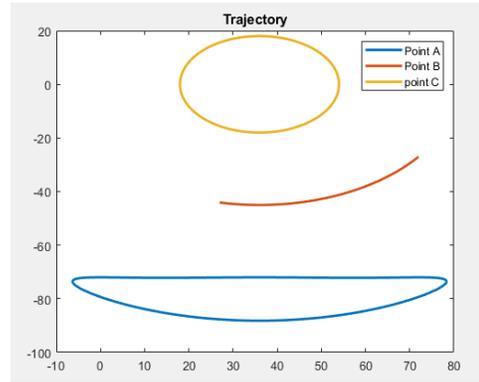

Figure 4. Trajectory

Figures 5,6 show how the angle and angle velocity of related variables change as the leg changes its q1, q2 and q3 angles. In one circulation, link 1 stably follows the desired kinematic parameters as q2 and q3 change themselves under the influence of q1.

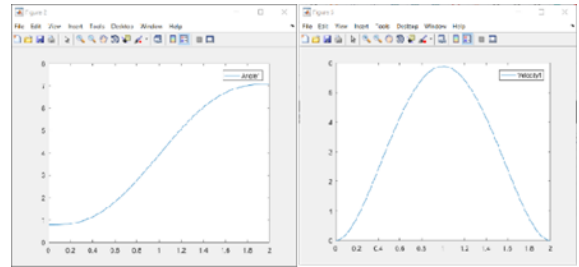

Figure 5. Angle 1 and Angle Velocity 1 Plots

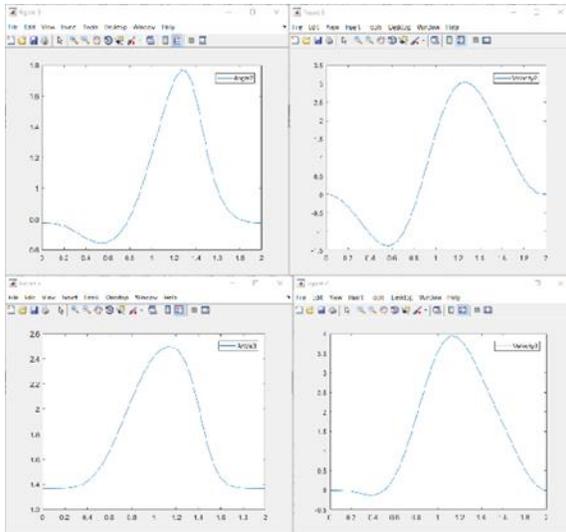

Figure 6. Angle 2,3 and Angle Velocity 2,3 Plots

Because the mechanism only has one degree of freedom, we only utilize one motor attached to one joint of link 1. The figure below shows the torque needed to be generated by the motor to drive the linkage system to reach desired dynamic configuration. The states before and after start position are not symmetric, therefore the torque at starting time is not the same as the one at ending time.

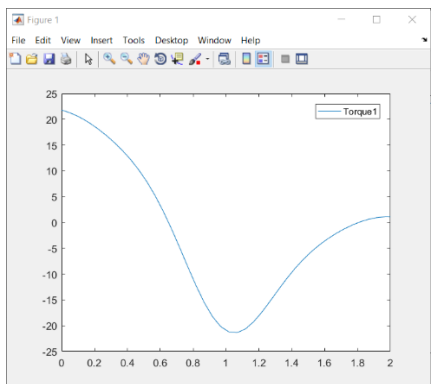

Figure 7. Torque 1 Plot

## Conclusion

This paper proposed a dynamic multi-linkage mechanism that can assist in human walking training. Simulations validated the motor's ability to support the mass of the mechanism itself and the mass of the patient's legs in all configurations. This can assist patients in tracking their leg's desired trajectory, thus one would ideally be able to perform more cycle rehabilitation practice repetitions while maintaining specific walking posture.

## Future Extension

For our device can not completely simulate the real situation of human walking, in our future work, we will collect more information about the trajectory related with human's hip, knee and ankle when they are walking, optimizing our model by changing the length of links or adding auxiliary links to design a more accurate trajectory suitable for real usage. Also, we will deeply smooth the trajectory to reduce the unnecessary torque damping between two circulations.

In terms of controller design, we will design a more powerful disturbance-observer-based-controller to compensate the disturbance from atmosphere, which can make virtual model simulation closer to real situation.

Y. Liu is a master's graduate student in The Fu Foundation School of Engineering and Applied Science, Mechanical Engineering, Columbia University, New York, NY 10027 USA.

Z. Yang is a master's graduate student in The Fu Foundation School of Engineering and Applied Science, Mechanical Engineering, Columbia University, New York, NY 10027 USA (email: zy2313@columbia.edu).